# Radiation Shielding Analysis for the PIP-II Linac at Fermilab    FERMILAB-CONF-22-709-AD

Igor Rakhno,* Nikolai Mokhov,# Igor Tropin,§ Sergei Striganov,Δ† Yury Eidelman†§

*Fermi National Accelerator Laboratory, P.O. Box 500, Batavia, Illinois 60510-5011, rakhno@fnal.gov
#Fermi National Accelerator Laboratory, P.O. Box 500, Batavia, Illinois 60510-5011, mokhov@fnal.gov
§Fermi National Accelerator Laboratory, P.O. Box 500, Batavia, Illinois 60510-5011, tropin@fnal.gov
†Fermi National Accelerator Laboratory, P.O. Box 500, Batavia, Illinois 60510-5011, strigano@fnal.gov
†§Euclid Techlabs, LLC, Solon, Ohio 44139, eidelyur@fnal.gov
ΔDeceased

## INTRODUCTION

The Proton Improvement Plan-II (PIP-II) [1] has been developed at Fermilab to provide powerful proton beams to the laboratory's experiments. An 800-MeV superconducting linear accelerator—a centerpiece of the project—is currently under construction in Batavia, Illinois (USA). After completion, the superconducting linac will be the starting point for the 1.2 MW (Phase 1) and 2.4-MW (Phase 2) proton beam that is needed for the Long-Baseline Neutrino Facility (LBNF) at Fermilab [2]. Due to unavoidable loss of a fraction of the beam in the accelerator components, a certain level of radiation will be generated in the accelerator tunnel both during normal operation and at accidents. This work deals with radiation shielding design for the accelerator facility.

## MARS15 Model of the Accelerator and Beam

A detailed computation model for the entire PIP-II Linac, Linac-to-Booster transfer line and corresponding shielding has been developed with the MARS15 Monte Carlo code [3]. Several parts of this model are shown in Figs. 1 thru 6. The accelerator model—based on engineering design and CAD geometry models—comprises major beam-line components including quadrupole and dipole magnets, solenoid magnets, superconducting accelerating cavities and cryomodules. Such a detailed model allows us to predict three-dimensional distributions of prompt and residual dose rate with a high level of accuracy. The MARS15 model is based on a built-in three-dimensional MAD-X based beam-line builder [4] and ROOT geometry [5] that provides great flexibility when building complicated geometry structures. Electromagnetic field distributions in the magnets and accelerating cavities were accounted for as well. Transport of charged secondary particles scattered back to an aperture of these elements is performed by means of the solvers comprised the ODEINT package of the BOOST C++ library. As an extra verification step, a comparison of individual trajectories has been made with analytical solution for energy gain along design trajectory of the linac [6] and trajectories generated by means of TraceWin code [7]. For energy gain on design trajectory, a comparison between our code and TraceWin code revealed perfect agreement. For distant trajectories, an acceptable agreement has been observed.

The accelerator shielding represents a bulk permanent structure with approximately a hundred of penetrations for both equipment and personnel. Current shielding design is based on an initial design developed during the initial stage of the project using conservative assumptions and simplified analytical methods. In fact, the presented shielding analysis is a shielding optimization study.

The PIP-II Linac is designed for negatively charged hydrogen ions (H$^-$) in order to mitigate space-charge effects inevitable for high power beams. At the end of the transfer line, two electrons will be stripped off each ion using the standard stripping foil technique, which ultimately produces a proton beam. In order to properly describe the beam transport in electromagnetic fields and beam interactions with matter, two new particles have been introduced to the MARS15 code, namely H$^-$ and H$^0$. Interactions of these H$^-$ and H$^0$ particles with matter are simulated using a model based on experimental data.

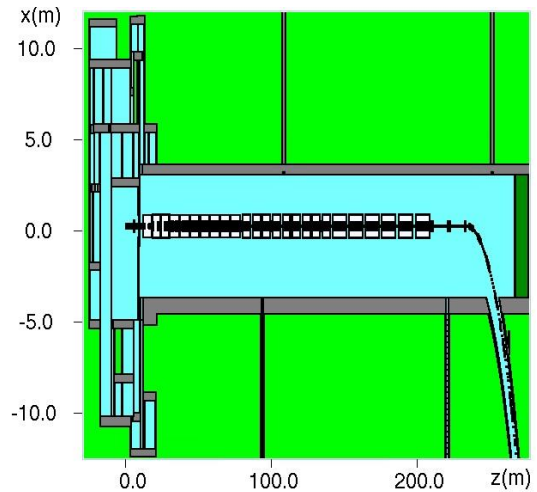

Fig. 1. A plan view of a major fragment of the model that shows the Linac with Front End Building and initial part of the Linac-to-Booster transfer line with shielding. The light blue, grey and light green colors correspond to the air, concrete and soil, respectively.

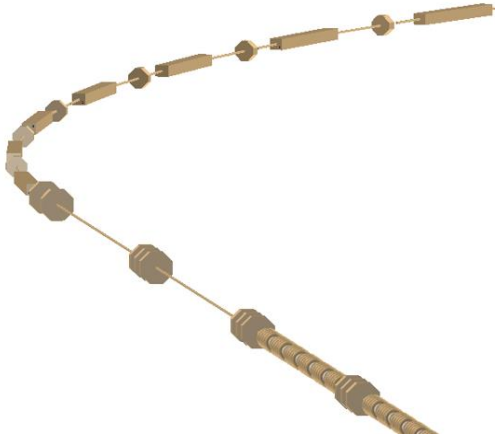

Fig. 2. A 3D view of the beamline components in the transition region from the Linac to the transfer line.

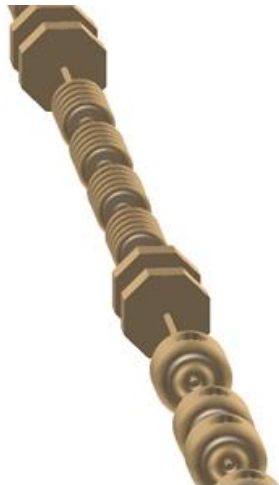

Fig. 3. A 3D view of several beamline components that belong to a single cryomodule.

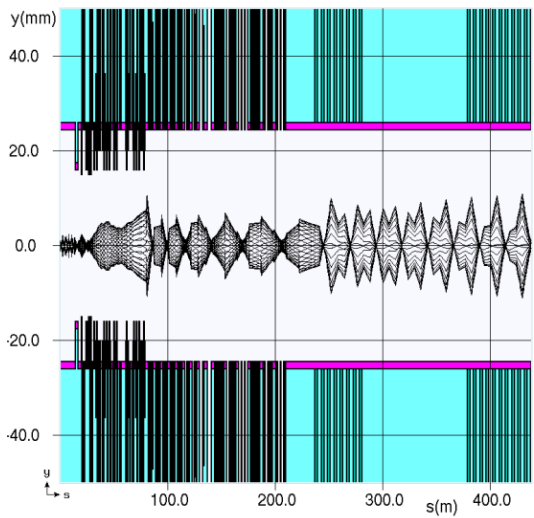

Fig. 4. A sample set of H$^-$ trajectories in the Linac. The region with s from 20 to 80 m contains beamline components with apertures less than 20 mm.

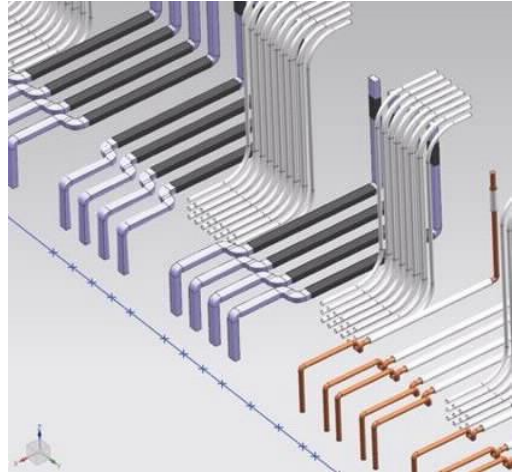

Fig. 5. A schematic engineering rendering of the RF waveguides and cable penetrations along the Linac.

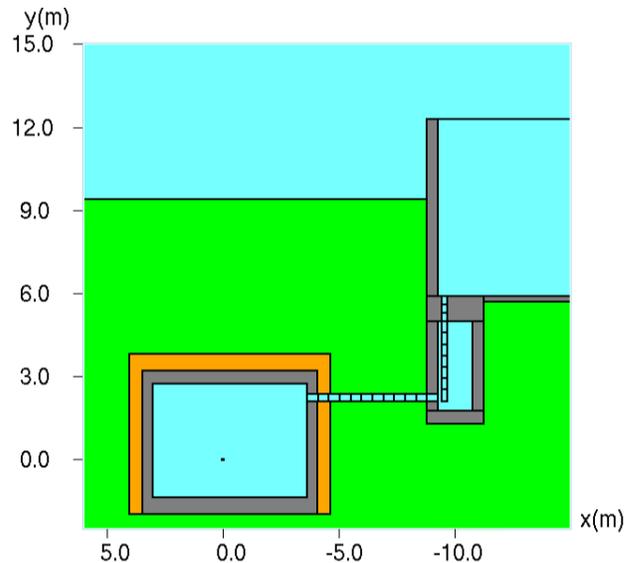

Fig. 6. A fragment of the model that shows a cross section of the Linac tunnel, klystron gallery and penetrations for the RF waveguides. The beige color corresponds to a crashed rock layer around the tunnel.

**Beam Loss**

A worldwide-spread *1 W/m rule* of uniform beam loss rate during normal operation, derived at the brainstorming workshop [8] from hands-on maintenance conditions for proton energy above 200 MeV, is used in this study as overall normalization in corresponding sections of the Linac. For normal operation, our goal is to make sure the prompt dose rate outside the Linac shielding does not exceed 0.5 µSv/hr.

As the worst-case beam accident scenario, we follow the approach used for the ESS linear accelerator [6]. In this case, the misbehaved beam of a full intensity is assumed to hit the beam pipe upstream of the corrector doublet in the last Linac section that corresponds to the highest particle energies. Duration of the accident is assumed to be 3 seconds and the angle of incidence is 2.5 milliradian. Our goal is to make sure that, due to the accident, prompt dose rate both atop the Linac shielding and in the klystron gallery will not exceed 0.01 mSv/hr.

**RESULTS**

Various prompt dose rate distributions have been calculated: along the Linac and Linac-to-Booster transfer line, on the berm and in the klystron gallery. A sophisticated combination of splitting and Russian roulette has been used in order to deal with the deep penetration problem (*i.e.* thick shielding above the accelerator tunnel). A comparison between this accurate approach and a simplified one mentioned above [6] confirmed that the latter represents a reasonable approximation to the accurate solution. Figures 7-8 and 9-10 show calculated dose distributions for normal operation and the accident scenario, respectively. It is worth mentioning that the exponential fitting works well not only for normal operation when relatively long flat regions can be present (see Fig. 7, z from 210 to 240 m), but for localized accidents as well (see Figs. 9-10).

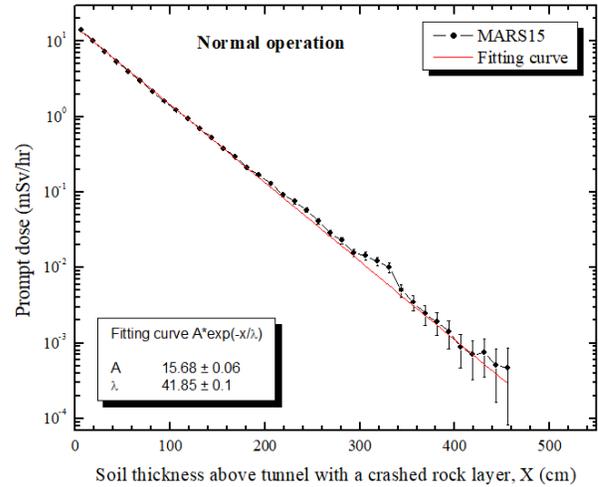

Fig. 8. The calculated prompt dose distribution above the Linac tunnel at normal operation, averaged along z axis from 210 to 240 m (see Fig. 7), and an exponential fitting function.

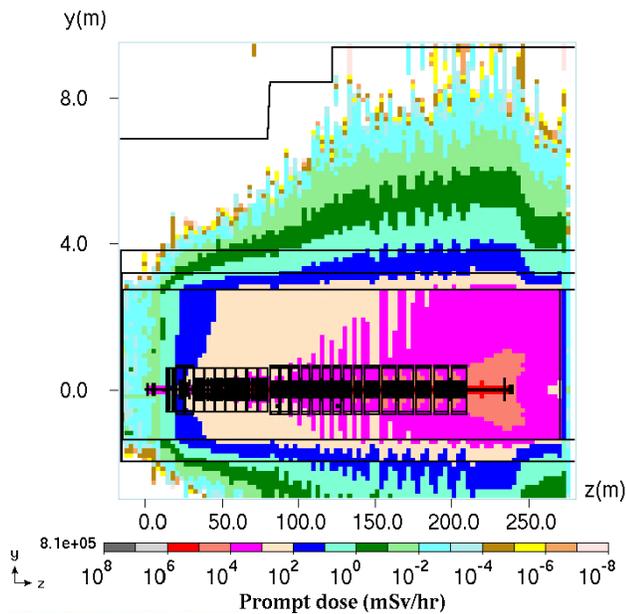

Fig. 7. A calculated distribution of prompt dose (elevation view) along the Linac at normal operation. The irregularities in the distribution along z axis are due to essentially heterogeneous structure of the beamline model introduced by cryomodules and accounting for electromagnetic fields in accelerating SRF cavities in the cryomodules.

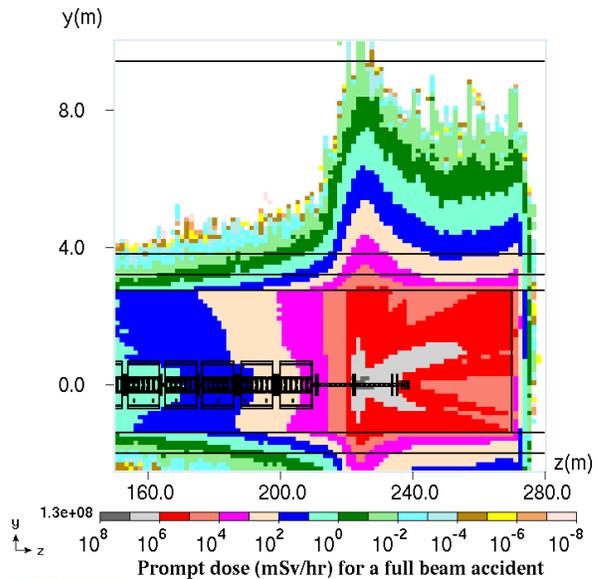

Fig. 9. A calculated distribution of prompt dose (elevation view) along the Linac at the accident assuming one accident per hour.

Prompt dose distributions along the multiple penetrations is a separate topic in this study. Calculations revealed that the round cable penetrations are much less important than larger rectangular penetrations for the RF waveguides (see Fig. 5) from the standpoint of enhanced radiation streaming. Also, detailed calculations revealed that the goal of not exceeding 0.01 mSv/hr in the klystron gallery at the accident can be achieved at a relatively modest price tag, namely using a concrete lid as thick as 90 cm in the RF vault (see Figs. 6, 11 and 12).

Other distributions, essential from the radiological standpoint, have been calculated as well: surface water activation, beam line component activation, residual dose, air activation as well as a skyshine contribution.

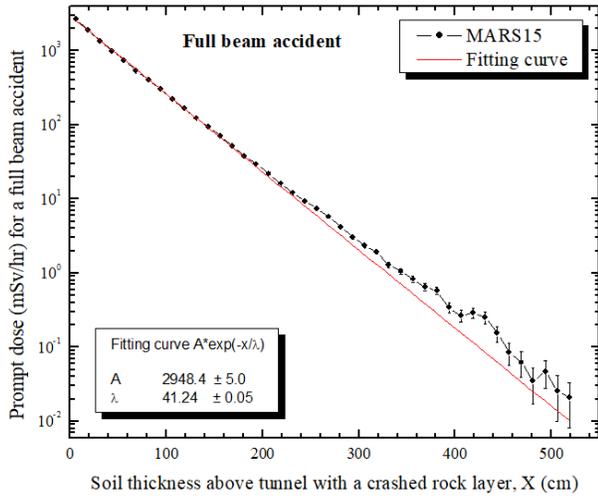

Fig. 10. The calculated prompt dose distribution above the Linac tunnel at the accident, averaged along z axis from 222 to 228 m (see Fig. 9), and an exponential fitting function.

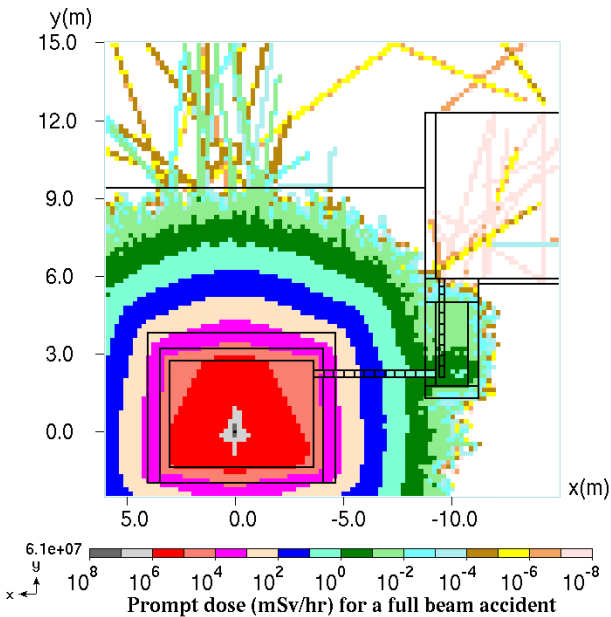

Fig. 11. A calculated distribution of prompt dose around the Linac (cross section at z = 227 m) at the accident assuming one accident per hour.

## ENDNOTES

This work is supported by Fermi Research Alliance, LLC under contract No. DE-AC02-07CH11359 with the U.S. Department of Energy.
This research used, in part, an ALCC allocation at the Argonne Leadership Computing Facility, which is a DOE Office of Science User Facility supported under Contract DE-AC02-06CH11357.

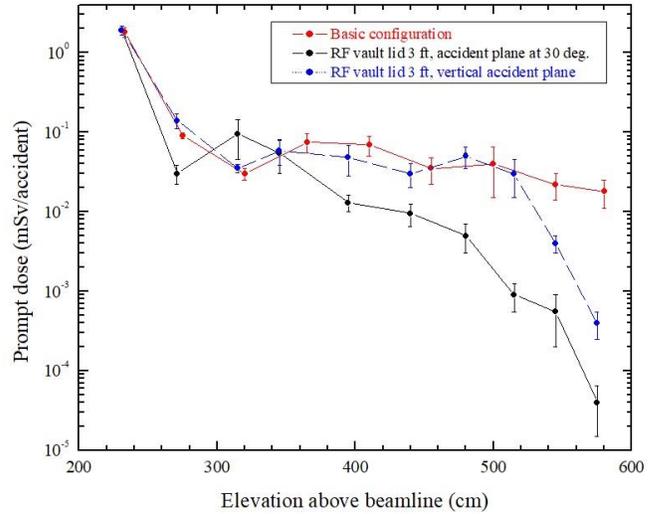

Fig. 12. The calculated prompt dose distribution in the RF vault (see Figs. 6 and 11). Basic configuration means using a lid as thick as 30 cm.